# A Wearable Virtual Touch System for Cars

Gowdham Prabhakar, Priyam Rajkhowa and Pradipta Biswas

Indian Institute of Science, Bangalore, India

pradipta@iisc.ac.in

**Abstract:**
In automotive domain, operation of secondary tasks like accessing infotainment system, adjusting air conditioning vents, and side mirrors distract drivers from driving. Though existing modalities like gesture and speech recognition systems facilitate undertaking secondary tasks by reducing duration of eyes off the road, those often require remembering a set of gestures or screen sequences. In this paper, we have proposed two different modalities for drivers to virtually touch the dashboard display using a laser tracker with a mechanical switch and an eye gaze switch. We compared performances of our proposed modalities against conventional touch modality in automotive environment by comparing pointing and selection times of representative secondary task and also analysed effect on driving performance in terms of deviation from lane, average speed, variation in perceived workload and system usability. We did not find significant difference in driving and pointing performance between laser tracking system and existing touchscreen system. Our result also showed that the driving and pointing performance of the virtual touch system with eye gaze switch was significantly better than the same with mechanical switch. We evaluated the efficacy of the proposed virtual touch system with eye gaze switch inside a real car and investigated acceptance of the system by professional drivers using qualitative research. The quantitative and qualitative studies indicated importance of using multimodal system inside car and highlighted several criteria for acceptance of new automotive user interface.

## 1.    Introduction

Early stages of automotive User Interface (UI) used analog interfaces consisting of buttons and switches, which require physical touch to manipulate. In physical touch based interfaces [Schnelle 2019], either the physical control needed to be visually searched or the feedback of selection needed to be visually confirmed. Although the visual search or feedback could be avoided with practice but it often distracted drivers from their visual attention in driving. As physical buttons cannot adapt to dynamically varying content, digital screens with GUI (Graphical User Interface) took over the market. These GUI-based systems accommodated new functionalities in addition to existing features and later gained popularity in the name of IVIS (In-Vehicle Infotainment System). Introduction of new functionalities in IVIS affects attention and learnability of drivers to interact. NHTSA reported 17% of car crashes due to operating such systems while driving [NHTSA 2012]. Though researchers are studying different input modalities, reducing cognitive load, visual, and physical effort to operate IVIS needs further investigation. This paper investigated the existing modalities of interaction to use such systems for undertaking secondary tasks and proposed new modalities involving virtual touch that requires limited effort minimising effect on driving performance. We proposed a new interactive device involving a laser tracker and eye gaze tracker for operating IVIS. Initially, we discussed the design and performance of a wearable laser tracker, finger movement tracker, and eye gaze tracker in different environmental conditions with varying ambient light and vibration. We evaluated the efficacy of laser tracker in laboratory as well as in-car environment, followed by user acceptance testing through qualitative research. The main contributions of this paper are as follows

1. Proposing a new multimodal user interface using virtual touch based pointing and eye gaze based selection

2. Evaluation of eye gaze tracking and wearable virtual touch interface in automotive environment

3. Designing a qualitative study for the automotive environment and undertaking thematic analysis to identify acceptance criteria for the new automotive user interface.

The rest of the paper is organised as follows. The next section presents literature review on virtual touch interfaces. Section 3 and 4 explain the design and setup of laser tracker and other virtual touch-based input modalities. Section 5 presents pilot studies for analysing the performance of the proposed system in different environmental conditions. Section 6 discusses the evaluation of the laser tracker in a driving simulation environment. Section 7 discusses the in-car study to evaluate the performance of laser tracker in real car environment. Section 8 investigates user acceptance of laser tracker system through qualitative analysis followed by conclusion in section 9.

## 2.    Related work

Kern (2009) classified five main interaction areas in a car – Windshield, Dashboard, Centre Stack, Steering Wheel, Floor and Periphery. Existing Graphical User Interface (GUI) based IVIS either has a Head-Down Display (HDD) at Centre Stack or Dashboard where the driver looks down to operate or a Head-Up Display (HUD) at Windshield, where the driver looks up to operate. They are operated using modalities like physical buttons, touchscreen, and voice recognition systems. Researchers have investigated hand gesture tracking [Ohn-Bar 2014, May 2014, Ahmad 2016, Prabhakar 2016], haptic feedback [Chang 2011, Vito 2019], personalizing displays to help drivers in parking vehicles [Feld 2013, Normark 2015]. In addition to these technologies, systems like virtual touch and intelligent interfaces are investigated by different research teams and described in detail in the following paragraphs.

### 2.1.    Virtual touch interface

Virtual touch systems aim to speed up the interaction with touchscreen system as they can activate a touchscreen without physically touching it. This paper investigated three types of technologies for virtual touch systems - infrared (IR), wearable device, and inertial measurement unit (IMU) based systems. The infrared sensors track the position of hand or fingers and have limitations in terms of field of view, accuracy, and latency of tracking in different vibrating and lighting conditions inside a car [Biswas 2017b]. The wearable systems using sensors attached to hand and fingers are also investigated for performing secondary tasks like wearable physiological sensors for music recommendation [Ayata 2018]. The third kind of virtual touch system is based on existing remote control-based pointers [Witkowski 2014], as illustrated by Witkowski. Khan has worked on a handheld device to interact with smart homes [Khan 2018]. Such systems are prone to unintended activation of functions in the infotainment system due to uncontrolled tracking of actions from fingers, hands, and remote even when the driver does not intend to operate the system.

## 2.2.    Intelligent interface

Pointing target prediction technologies were already investigated for human-computer interaction [Murata 1998, Lank 2007, Ziebart 2010] and, more recently for automotive user interfaces. Ahmad [2014] proposed a system that predicts the intended icons on the interactive display on the dashboard early in the pointing gesture. He reported improved performance of the pointing task by reducing the target selection time using a particle filter-based target prediction system with which the user was able to select icons on screen before his hand physically touched the screen [Ahmad 2016]. They also reported a reduction in workload, effort, and duration of completing on-screen selection tasks. Biswas [2013; 2014] reported neural network-based prediction algorithms to locate intended target, while Lank and Phillip [Lank 2007, Pasqual 2014] proposed a method to predict endpoint using motion kinematics. Biswas [2017b] also presented an intelligent finger tracking system to operate secondary tasks in cars.

## 2.3.    Wearable devices in automotive

Wearable devices are investigated in multiple disciplines of computer science like ubiquitous computing, Internet of Things (IoT), Assistive and Ambient Technology, and so on. Steinberger identified requirements for wearable devices in automotive as robust to body movement and capable of real-time data streaming [Steinberger 2017]. The data collection systems should be robust to movement artefacts associated with driving as it increases noise in data [Stern 2001, Baguley 2016].  Though reliable physiological measurement devices are heavyweight to wear and not suitable for automotive consumers [Liang 2007], advanced physiological measures are embedded in wearable devices like fitness trackers and smartwatches. Fitness trackers are capable of measuring heart rate activity and provide information regarding the user's seating status. Smartwatches measure biometric data, which is an indicator for driver drowsiness [Aguilar 2015]. Patterns are recognised from these devices' data to determine and characterise driver skills [Zhang 2010]. Wearable devices are used for acquisition of driver arousal data as an indicator of task engagement [Yerkes 1908]. Harman Becker automotive systems patented a head position monitoring device using a wearable loudspeaker that is worn on upper part of the body and a distance away from ears [Woelfl 2020]. Though consumers hesitate to use wearable technologies while driving, they still use wearable technologies for personal assistance like fitness wristbands [Kundinger 2020] which establishes a new market for acceptance of wearable technologies in everyday lives.

## 2.4.    Non-visual feedback

Researchers attempted to eliminate or reduce visual search using gesture recognition techniques, but the systems either require to remember a set of gestures (AirGesture System [May 2014]) or relative positions of screen items (BullsEye system [Weinberg 2012]). Additionally, such systems worked inferior to a touchscreen system in terms of driving performance or secondary task. It may be noted that, systems and services developed for elderly or disabled people often finds useful applications for their able bodied counterparts – a few examples are mobile amplification control, which was originally developed for people with hearing problem but helpful in noisy environment, audio cassette version of books originally developed for blind people, standard of subtitling in television for deaf users and so on. Considering these facts, technologies developed for users with visual impairment can have potential for non-visual interaction with IVIS system. A plethora of research has been conducted on navigation applications for blind users [Ganz 2011, Mulloni 2011, Rocha 2020]. Gorlewich [2020 ] and Palani [2020] formulated a set of guidelines for haptic and tactile rendering of touchscreen elements while ISO/TC 159/SC 4/WG 9 is working on a general purpose ISO standard on haptic and tactile interaction. It will be challenging to design an IVIS with 4mm inter-element spacing based on Gorlewich's [2020] study, however, it may also be noted that Prabhakar [2020] designed an automotive Head Up Display with a minimum inter-element spacing to make it accessible through eye gaze and gesture-based interaction and the interactive HUD improved driving performance with respect to traditional touchscreen in driving simulation study.

## 2.5. Summary and proposed approach

Existing research mostly explored infrared-based sensors in automotive environment and least considered other options offering virtual touch. Wearable technologies are mostly explored for physiological measurement but not as a direct controller of user interfaces inside car. The tactile and haptic feedback systems are promising but needs more investigation for integration to automotive environment and display versatile content of an existing IVIS.

In this paper, we proposed a new virtual touch device using a laser tracker for operating secondary tasks in a car. We have discussed different stages of development and evaluation of the laser tracker for operating any display (GUI) without physically touching the display. We evaluated three different modalities for undertaking pointing and selection tasks in a GUI and chose the best performing modality (laser tracker) to be assessed in an automotive environment. We tested the robustness of the laser tracker as well as eye gaze tracker under different luminance conditions. We integrated this system in an automotive environment (driving simulator) and evaluated its performance with that of the existing touch screen displays for operating secondary task (dashboard display). We investigated factors influencing user acceptance of our proposed system using qualitative analysis.

# 3. Design of laser pointer based virtual touch system

Traditional laser pointers are used for pointing on a projected screen. However, operating a GUI also involves selecting a target. Designing a laser tracking based virtual touch system addressed the following two challenges

- Designing a selection mechanism – in particular, mechanical switch and eye gaze tracking based selection mechanisms were explored
- Designing a software module to accurately detect a screen element which the user is pointing at it

In this section, we discussed the design and development of laser tracker with mechanical switch as well as eye gaze switch. We have also discussed the development of touch switch on the steering wheel for automatically turning the laser module OFF while the user does not intend to undertake any secondary task.

## 3.1.    Laser tracker with mechanical switch

The laser tracker console is constructed using a laser module and three button switches, as illustrated in Figure 1. The laser module and switches are connected to a microcontroller board (Arduino Uno) [Prabhakar 2016]. Each switch is used to perform left, right, and double-click operations like that of a computer mouse. The working of the laser tracker system is explained in further detail as follows.

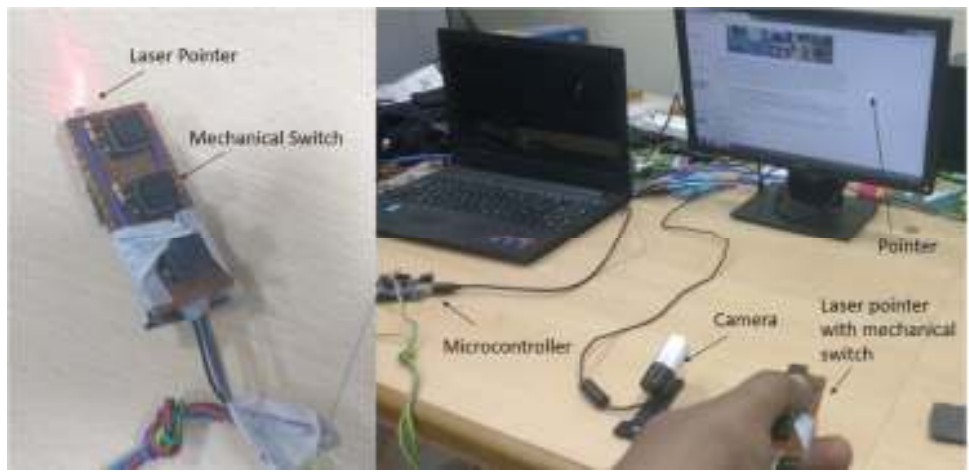

Figure 1. Laser tracker with mechanical switch system

## 3.2.    Laser tracker with eye gaze switch

We also proposed a technique where we use an eye gaze tracker as a switch to trigger the operation of selecting the target pointed by a laser pointer. In this technique, we initiate a trigger when we find the user looking at a location within a viewing angle of 1.6° for 300ms. The optimal boundaries of viewing angle and glance duration are determined by conducting a study, as discussed in section 5.3.

This wearable device has a feature to turn the laser ON and OFF. In addition to eye gaze switch, we can also select the target on the screen using a thumb tap (optional modality of target selection). This wearable module consists of two parts. The first part facilitates the function of selecting a target on the screen by tapping thumb with index finger. The second part facilitates the function of turning the laser OFF while the user's hand is on the steering wheel and ON while the user's hand is off the steering wheel.

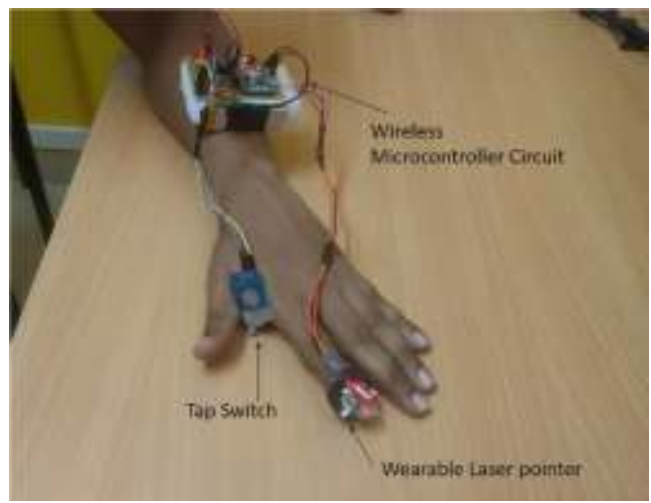

Figure 2. Wearable laser tracker with touch (tap) sensor

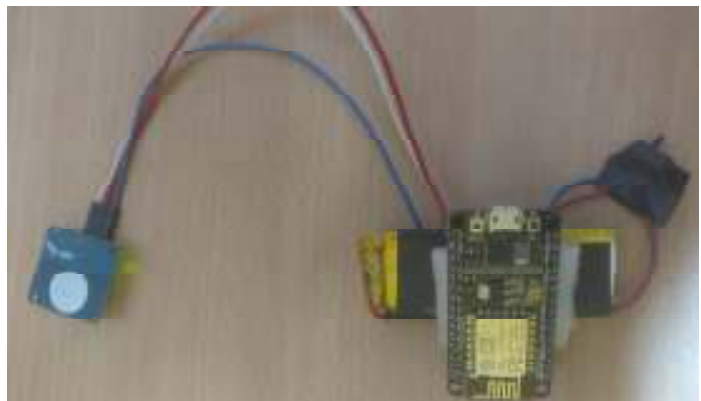

Figure 3. A touch sensor to be attached to the steering wheel

## 3.3.    Integration to driving simulator

The laser tracker system requires a physical setup of a camera facing the IVIS without obstructing the participant. In this section, we discussed the setup of driving simulator and integration of the laser tracker system.

### 3.3.1.    Driving simulator

We used a Logitech steering wheel, pedals, and a 43" LCD (Liquid Crystal Display) screen along with the ISO 26022 Lane Changing Task (LCT) [Mattes 2003]. The lane changing task involved driving in any one of the three different lanes in the simulator and changing lanes at a random interval. The lane-changing signboards were shown alongside the road. For each lane change, there was a reference path along which the driver was intended to move. The task did not involve any other traffic participant (vehicle, pedestrian etc.) on the road. The driving performance was measured in terms of deviation from the designated lane, steering angle, and average speed.

### 3.3.2.    Dashboard display

Our user studies involved participants in undertaking the lane changing task and simultaneously undertake a secondary task in terms of a pointing and selection task in a dashboard display. A touchscreen display (Figure 4) was used to simulate the dashboard display for secondary task in driving simulator. Participants were asked to select the targeted button when they heard an auditory cue. The target button was indicated by change of colour from light grey to bright yellow. The time taken to select every button was recorded.

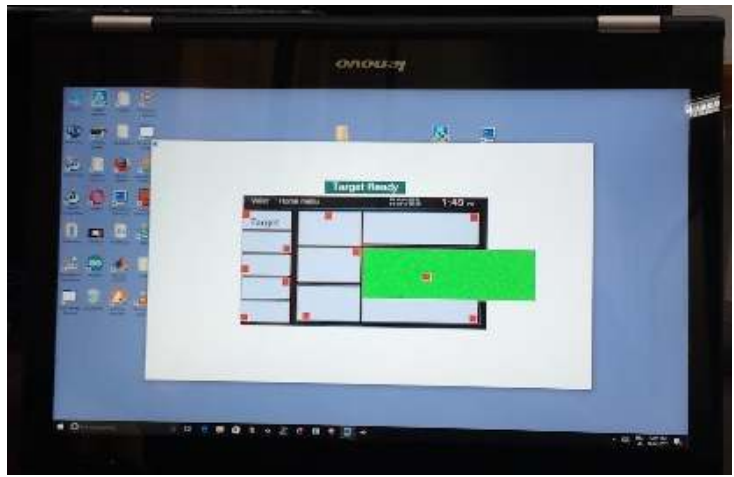

Figure 4. Setup of dashboard display in a laptop screen

### 3.3.3.   Laser tracker

The laser tracker with mechanical switch (Figure 5) and laser tracker with eye gaze switch (Figure 6) were used by participants to select target buttons on IVIS display while driving.

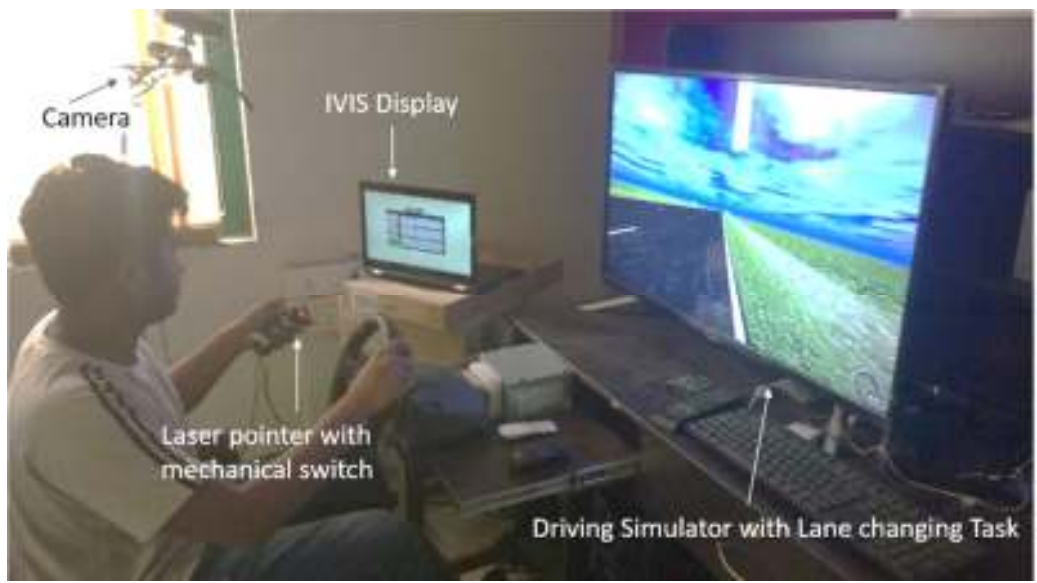

Figure 5. Participant performing a secondary task using laser tracker with a mechanical switch

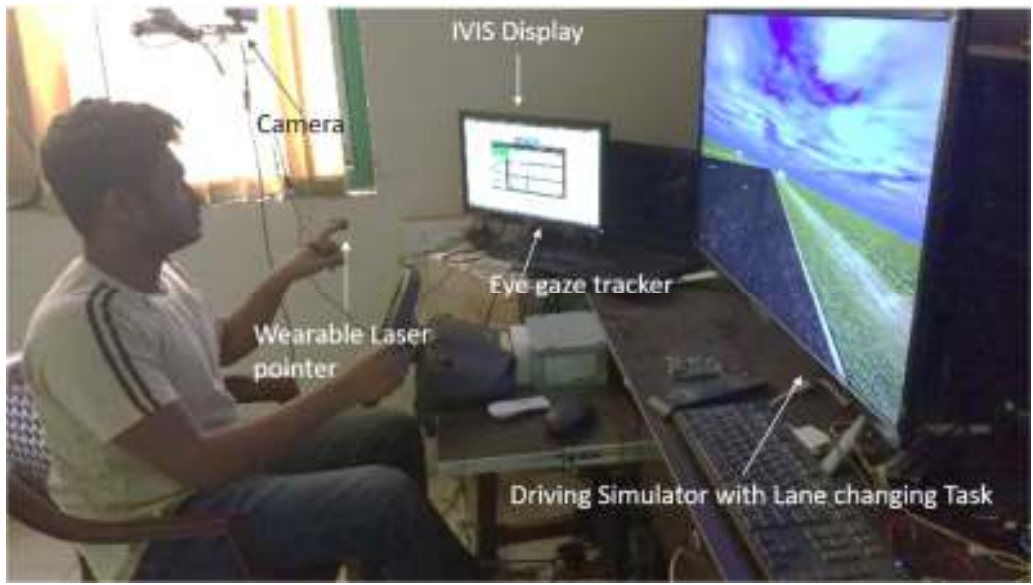

Figure 6. Participant performing a secondary task using laser tracker with eye gaze switch

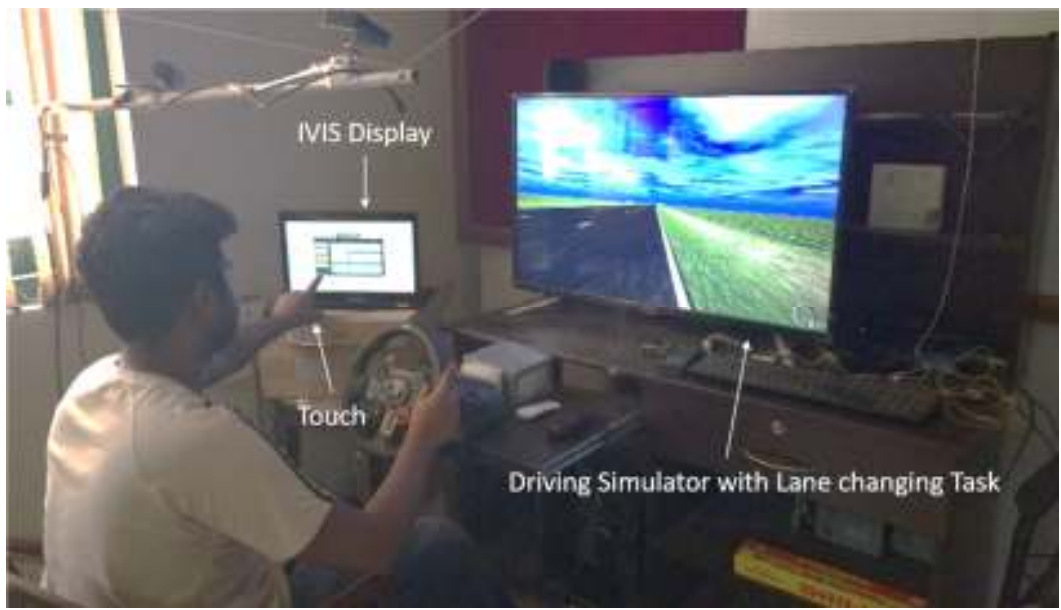

Figure 7. Participant performing a secondary task using touch-input modality

## 4.   Other input modalities

In this section, we presented two different alternative technologies to implement virtual touch system those were compared with the laser tracker based virtual touch system.

We also described the touch based system that was used as a baseline in subsequent quantitative user studies.

## 4.1. Touch-input modality

In the existing infotainment systems of cars, touch-input is the widely used modality of interaction. In this modality, the user is intended to select different targets on the display by physically touching the display. As soon as the intended target is touched by the driver, the object located in the approximated touched coordinates is selected. This modality requires the driver to gaze at the dashboard and perform a motor action to select the target. This system is illustrated in Figure 7.

## 4.2. IMU tracker

IMU is an electronic device that measures its orientation in 3D space. Since the IMU (Xsens MTi-1s) is made up of a 3-axis gyroscope, accelerometer, and magnetometer, the fused data from these sensors are used to obtain the Roll, Pitch, and Yaw angles. The system worked by attaching the IMU sensor to a pair of fingers in the user's hand and controlling the mouse cursor movement in accordance with the finger-pair movement. We mapped the position angles to the coordinates on the display screen. Since the operation was done on a 2-D display screen, only the pitch and yaw angles were considered for determination of x and y coordinates on the display screen. The pitch angle (pitch) was mapped to the y-coordinate (ycoord), and the yaw angle (yaw) was mapped to the x-coordinate (xcoord). A click was made if the cursor dwelled at a fixed position for 1.5s. The mapping was done using equations (1) and (2), where LL denotes left limit, RL denotes right limit of the yaw angle, and BL denotes bottom limit, and TL denotes Top limit of pitch angle. Variables *resx* an *resy* denote screen resolution in pixels.

$$xcoord = (LL - yaw) \times \frac{resx}{|RL - LL|} \qquad (1)$$

$$ycoord = (BL + pitch) \times \frac{resy}{|TL - BL|} \qquad (2)$$

The position angles vary as the orientation of the IMU sensor varies. So, limits for the pitch (BL and TL) and yaw angles (LL and RL) were decided by the calibration process. In the process of calibration, the user was instructed to move the finger pair attached with the sensor to corners of the display screen. The maximum and minimum limits of each angle were obtained in this process using equations (3), (4), (5), and (6).

Let (x1, y1), (x2, y2), (x3, y3), (x4, y4) be the four coordinates of the four corners of the display screen during calibration. Then,

$$Pitch\ TL = min(y2, y1) \tag{3}$$

$$Pitch\ BL = max(y3, y4) \tag{4}$$

$$Yaw\ LL = max(x1, x4) \tag{5}$$

$$Yaw\ LL = min(x2, x3) \tag{6}$$

## 4.3. IR based tracker (LeapMotion)

The LeapMotion sensor (LM-010), consisting of two monochromatic IR cameras and three infrared LEDs (Light Emitting Diode), was used to track the finger positions in a 3D coordinate system defined by itself. The equation of the 2D screen in its 3D coordinate system was evaluated, and the following set of equations (7) and (8) was used to take an orthogonal projection of finger position on the screen.

$$ScreenX = \frac{ScreenWidt}{w} \times (ftp.x + a) \tag{7}$$

$$ScreenY = \frac{ScreenHeig\ t}{} \times (b + c \times ftp.y - d \times ftp.z) \tag{8}$$

Constants *a*, *b*, *c*, *d*, *w*, and *h* were calculated based on the relative screen position with respect to the LeapMotion sensor, and *ftp* refers to the Fingertip Position. The mouse cursor was moved in accordance with coordinates corresponding to this finger position on the 2D screen. When the cursor dwelled at a fixed position for 1000ms, a left-click was invoked at that coordinate.

### 4.4.    Target adaptation algorithm

Pointing tasks are traditionally modelled as a rapid aiming movement. Woodsworth first proposed the idea of the existence of two phases of movements in a rapid aiming movement, ballistic main movement phase and homing phase [Woodworth 1899]. Once a pointing movement is in the homing phase, we can assume the user is near to his intended target. We used a simple target expansion algorithm in ISO pointing task to reduce pointing and movement time by detecting the homing phase of a cursor movement. As the cursor moves on the screen, we continuously measured the instantaneous velocity and acceleration of the cursor. If the acceleration was negative or the velocity was zero, we expanded the nearest target to the cursor's location to 1.5 times its original size.

## 5.    Quantitative user studies

We undertook a series of quantitative user studies to measure and compare the performance of the laser tracker system with respect to existing technologies. Table 1 below summarises all quantitative studies reported afterward.

**Table 1.** Summary of Quantitative User Studies

| Study No. | Environment | Modalities of Interaction | Aim |
|---|---|---|---|
| 1 | Laboratory | Laser tracker, IR, and IMU-based finger trackers | Comparing response times and driving performance with secondary task undertaken by laser tracker, IR, and IMU-based finger trackers |

| 2 | Laboratory | Laser tracker | Optimising laser tracker for different lighting conditions |
|---|---|---|---|
| 3 | Outdoor | Eye gaze tracker | Comparing the performance of eye gaze controlled interface in different lighting conditions |
| 4 | Laboratory | Eye gaze tracker | Analyse for the average glance duration and visual angle during selection task. |
| 5 | Simulator | Laser tracker, eye gaze tracker, touchscreen | Comparison of the performance of laser tracker in terms of average selection time and driving performance with touchscreen |
| 6 | Inside Vehicle | Laser tracker, eye gaze tracker, touchscreen | Evaluation of the efficacy of laser tracker in terms of average selection time with touchscreen inside a vehicle |

We conducted pilot studies to investigate parameters like ambient light, vibration, colour space, and camera exposure for evaluating the efficacy of trackers under different environmental conditions inside cars. We discussed our pilot studies in the following sections.

## 5.1.　Comparing laser tracker with LeapMotion and IMU

Pointing devices for Graphical User Interfaces (GUI) are conventionally evaluated using ISO 9241 pointing task. This study focused on different sensors tracking wrist and finger movements and proposed algorithms to control an on-screen pointer using those devices [Prabhakar 2016]. We used a laser tracker, Xsens IMU tracker, and

**15**

LeapMotion tracker as cursor controller and evaluated their performance using ISO 9241 pointing task. We also evaluated an intelligent algorithm that predicts pointing targets and helps to reduce pointing and selection times. We compared Indices of Performance (IP) using Fitts' law [Fitts 1954] in equation (10), selection time, and mental workload for different pointing devices.

**Participants:** A total of 9 individuals participated in the evaluation task. Participants were young students from the campus with an average age of 26 years (2 female, 7 male).

**Material:** Tasks were performed on a Windows PC running Windows 7 on 2.0 GHz Intel Core-i3 processor and 4GB RAM. The size of the display screen was fixed at 435 mm × 325mm with a resolution of 1024×768 pixels. The size of the laser point was 3.5mm in diameter on the display screen. As the user was only performing the pointing task looking at the screen, we used the laser pointer with mechanical switch. The list below shows different sizes of targets and distances of their separation from the centre of the display.

- Width (W) of targets: 45, 55, 65, 75
- Distances from center (D): 80, 160, 240, 325
- Pixel width: 0.42mm per pixel

**Design:** A standard procedure for target selection was executed to evaluate different modalities. In this procedure, the participant was instructed to perform a pointing task [Biswas 2015] which was like ISO 9241 pointing task. It contained the target in white colour and the obstacles/distracters in blue colour, as shown in Figure 8. The cursor was brought to the centre of the screen before each target was clicked to note down the distance and time taken for the movement of cursor from the centre of the screen. The participant was instructed to navigate the cursor through the screen and select the target using each modality. Each modality was evaluated both in regular as well as adaptive modes. In adaptive mode, the size of the target, as well as distracters, vary depending on the cursor movement. The average time taken for a participant to move the cursor from the centre of the screen to the target was calculated for each modality. Then the Index of Difficulty (ID) [Fitts' 1954] was calculated by using equation (9).

$$ID = log_2\left(\frac{2D}{W}\right) \tag{9}$$

$$T = a + blog_2\left(\frac{2D}{W}\right) \tag{10}$$

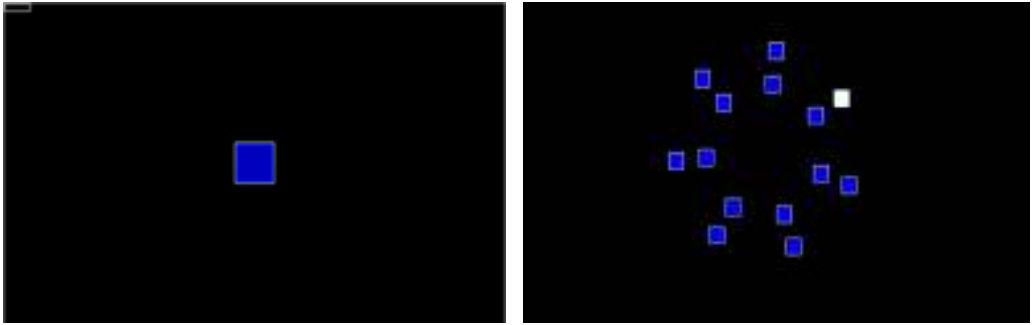

Figure 8. Screenshots of the center selection (left) and target selection (right)

**Procedure:** Two pointing tasks were performed using each modality in which the former was conducted in normal mode, and the latter was conducted in adaptive mode. A total of 3(modalities) × 2 trials were performed by each participant. All trials were randomised to avoid the order effect. Each trial was conducted for about 4 minutes logging 120 target hits per participant. The corresponding mental workload of participants was also recorded using NASA's TLX (Task Load Index) and SUS (System Usability Scale) scores for each trial.

### 5.1.1. Results

In the following bar charts, heights of bars represent arithmetic mean, and the error bars represent standard deviation.

**Pointing and Selection Time:** We compared the average pointing and selection times for different combinations of target width and distances. Figure 9 shows the graph of selection times with respect to IDs for different modalities. We found that participants took the least selection time using the laser tracker while highest using the LeapMotion controller. It is observed from Figure 9, that the laser tracker takes an average time of 1315ms for regular pointing and selection task. The adaptive modality of laser tracker further reduced the time of selection down to 1156ms. The LeapMotion tracker took an average time of 1732ms for regular task and 1703ms for adaptive task. The IMU tracker took an average time of 2181ms for regular task and 2115ms for adaptive task. A repeated measure ANOVA on selection time showed significant difference between modalities [$F(2,90) = 75.1$, $p < 0.01$]. A pairwise t-test on selection times of laser and IMU trackers showed significant difference ($p<0.05$) between adapted and regular modes.

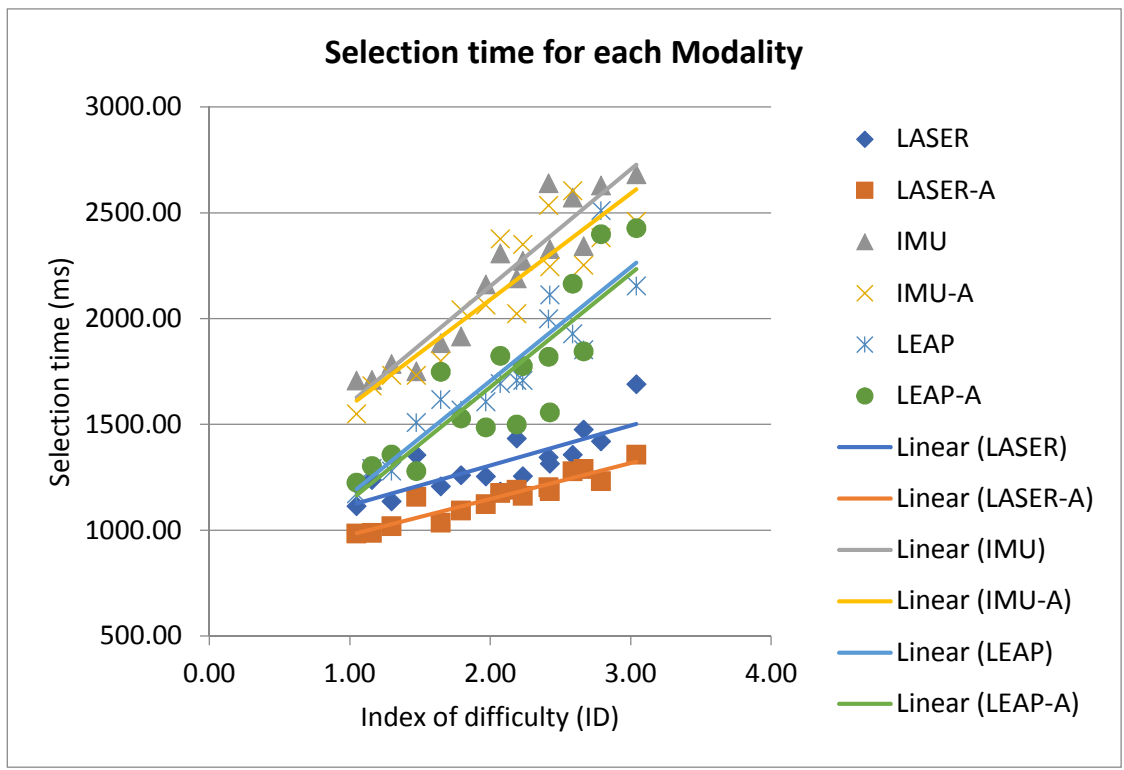

Figure 9. Comparing pointing and selection times for different modalities (-A stands for Adaptive mode)

Figure 10 shows the Index of Performance (IP) for all pointing devices. The highest IP was found for the laser tracker.

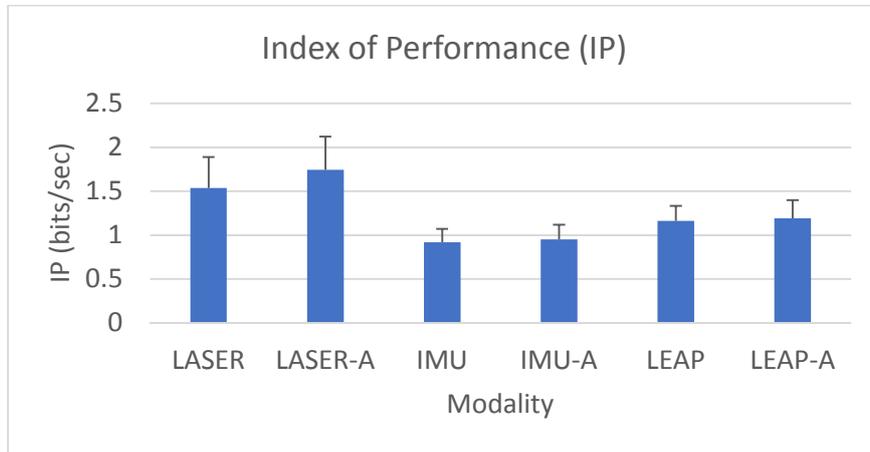

Figure 10. Comparing IP for different modalities

**Subjective Feedback:** The mental workload was highest for the LeapMotion controller and least for laser tracker (Figure 11). Differences among TLX scores were not significant in a one-way ANOVA.

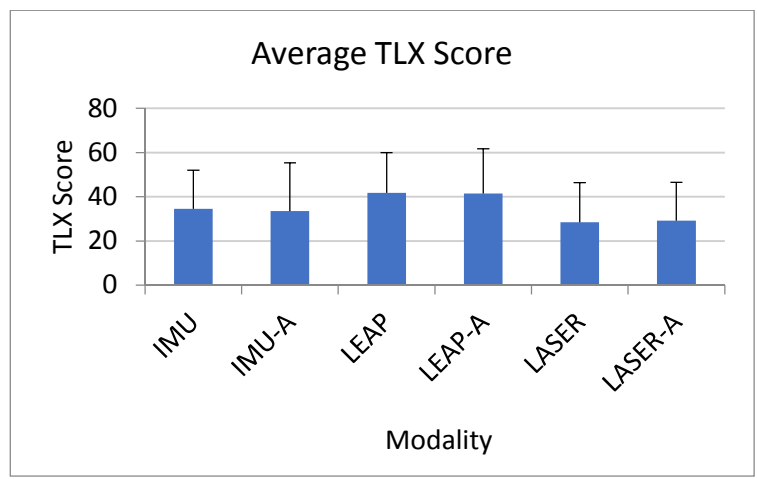

Figure 11. Comparing mental workload in terms of TLX scores

The average SUS scores were calculated for each modality, as shown in Figure 12. All SUS scores were above 68, indicating that participants preferred all modalities. The highest score was noted for laser tracker, while the lowest was found for the LeapMotion controller.

**Wrong Selection:** Users clicked wrong objects/distracters several times while performing trials that were not accounted for calculating average selection time. These clicks are referred to as wrong selections.

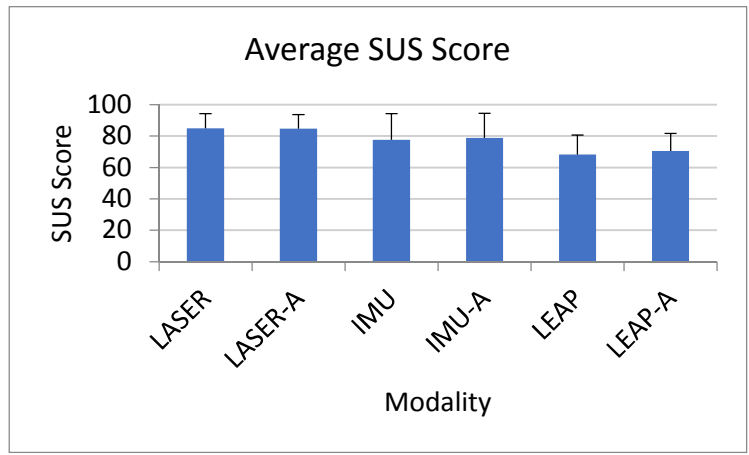

Figure 12. Comparing subjective preference in terms of SUS scores

The graph in Figure 13 showed that the LeapMotion controller had more wrong selections. It was also found that the number of wrong selections was more in adaptive tasks in every modality.

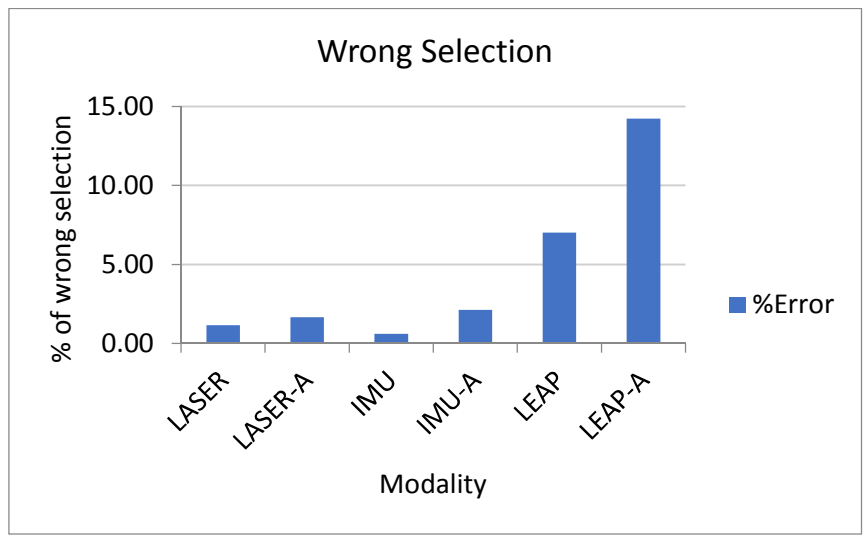

Figure 13. Comparing the percentage of wrong selection

## 5.1.2. Discussion

Existing gaming, automotive and military aviation environments explored finger or hand movement tracking-based interaction as a gesture recognition system. However, any gesture recognition system requires users to remember a set of gestures like commands of early-stage command prompt system. We explored new pointing modalities where an on-screen pointer can directly be controlled by finger or wrist movement. It is different from existing hand or air-gesture systems because users need not to remember any specific gesture and can control an on-screen pointer by small finger movement in the right, left, up, and down directions.

We proposed algorithms to control the cursor using three different types of sensors, as listed below

- Laser tracker uses a projected beam of light and uses an image processing technique to detect the position of the projected beam and control the cursor.

- IMU tracker uses inertial sensors to track the position of hand and finger and control cursor.

- LeapMotion controller uses infrared cameras to take videos of hand movement and control the cursor using the position of the index fingertip.

We also evaluated an adaptation algorithm based on the enlargement of targets to reduce pointing and selection times. A standard ISO 9241 pointing task found that the laser tracker is the fastest to operate, and the adaptation system can significantly reduce pointing and selection time. Users' mental workload and subjective preference also supported the result.

The IMU tracker had a default of 1.5s as dwell time for selection of the target. Reducing this delay will bring down the selection time of the target. The LeapMotion tracker required a continuous lift of wrist over the sensor, which might have contributed to its poor performance. The dwell time for selection in the LeapMotion controller was found to be too low to trigger wrong selections and was much worse in adaptive mode.

The result of this study showed that the laser tracker gave promising performance compared to other modalities. We chose laser tracker for integrating into an automotive environment and evaluated the efficacy of eye gaze tracker in different lighting conditions, as discussed in the following section.

## 5.2. Evaluation of eye gaze tracker in different lighting conditions

We conducted a user study to evaluate the efficacy of the existing eye tracker (Tobii 4c) under different lighting conditions: dark, light, and sun, where dark was the least and sun being the highest in luminance. As the tracker fails to detect eyes beyond a luminance value, the tracker was tested from the darkest to the brightest luminance outdoor under sunlight exposure to find the upper bounds of detection of eyes by the tracker. We used ISO 9241 pointing task described in previous section 5.1 to compare movement time with respect to different indices of difficulty. All trials were randomised to avoid the order effect. The ID vs. selection time of targets at different luminance conditions is shown in Figure 14.

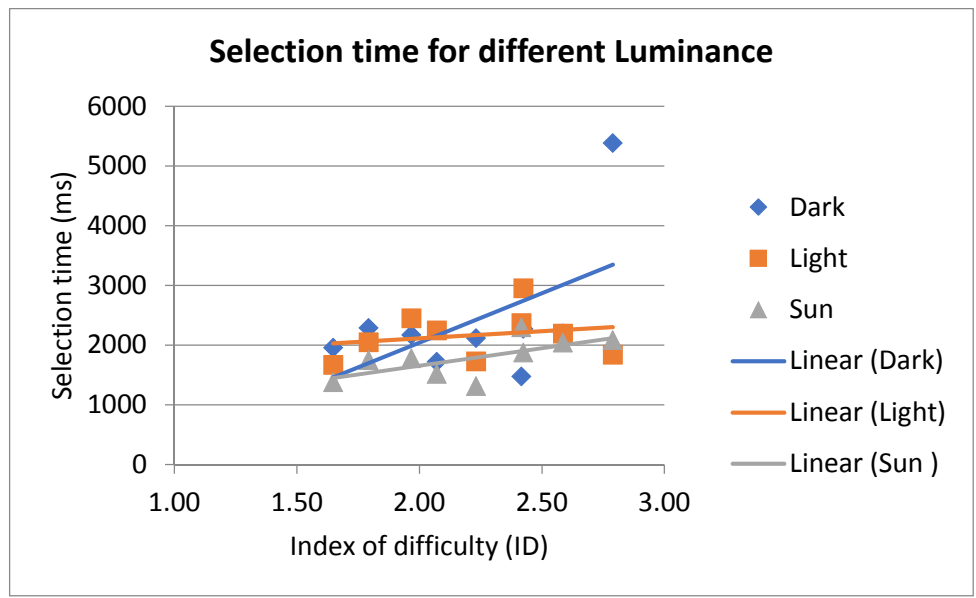

Figure 14. Average selection time for different luminance

In a pilot study involving 7 participants (4 male, 3 female, age range 23 to 35 years), we did not find a significant difference in performance in terms of pointing and selection times in a range of 6 to 1500 lux of illuminance. As we place the eye gaze tracker in the car's dashboard, we do not expect direct sunlight to be more than 1500 lux to hit on the dashboard. Hence, we conclude that change in lighting conditions in day and night would not affect the performance of gaze-controlled interfaces. As we used eye gaze tracker as a switch for the laser tracker system, we were required to estimate the average dwell time for locking the gaze position. In that direction, we conducted a pilot study to estimate the average dwell time of gaze-controlled interface, as discussed in the following section.

## 5.3. Eye gaze tracker dwell-time calibration study

We analysed an existing dataset for dwell time and average standard deviation of the eye gaze to find the visual angle boundary for performing a secondary task on an IVIS. The dataset had the eye gaze and timestamp information while performing a secondary task by touching the screen and selecting targets while driving. The user was instructed to select the icon on a touch screen IVIS highlighted as 'Target' after he/she hears an auditory cue. During selection, the eye gaze movement was tracked using an eye gaze tracker, and the time taken to complete the target selection task was recorded. Data was analysed for the average glance duration and average visual angle during selection.

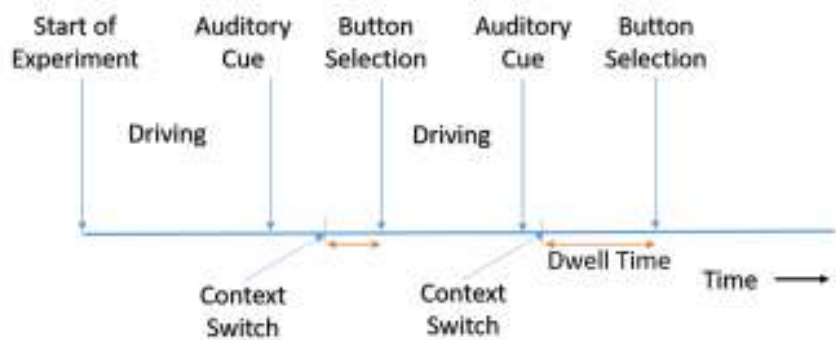

Figure 15. The timing diagram of the secondary task

The flow of the experiment is illustrated in the timing diagram (Figure 15). When the user looked at IVIS, the eye tracker detected eyes and recorded the gaze data. The context switch illustrated in the timing diagram explains the time taken by the user to switch his mind from driving to operating secondary task.

We found that the average X offset was between 100 and 200 pixels. The average Y offset was between 400 and 800 pixels. The average reaction time was between 800 and 1400ms, and the average glance duration was between 200 and 1000ms. These values for dwell time and visual angle in the existing algorithm were expected to improve the performance of the laser tracker with eye gaze switch system. We integrated the laser tracker system into the driving simulator. We conducted a user study to evaluate the efficacy of our system in comparison to the touch screen interface, as discussed in the following section.

## 6.    User study within driving simulator

Following the results of the pilot studies, we conducted a user study to evaluate the performance of laser tracker in the automotive environment. We compared the performances of laser tracker with mechanical switch, laser tracker with eye gaze switch and touch input systems for operating secondary tasks while driving.

### 6.1.    Participants

We recruited 10 participants (8 male and 2 female) with an average age of 26 years from our university for this study. All participants had at least a year old driving license and drove regularly. We let participants freely drive the simulator before data collection and made sure that the simulator experience did not influence the increase in task difficulty.

## 6.2. Material

A windows PC was used to run the laser tracker system. A Logitech G29 driving wheel was used for driving simulator running ISO 26022 lane changing task. For undertaking secondary task, we used the same set of materials described in section 3.

## 6.3. Design

In this dual-task study, participants were instructed to drive along a 3-lane motorway and undertake a lane-changing task. They were instructed to select targeted buttons on the dashboard display after they hear an auditory cue at regular intervals of 5 to 7 seconds. The following modalities were compared

- Laser tracker with mechanical switch (Laser_Mech)
- Laser tracker with eye gaze switch (Laser_ET)
- Touch (Touch)

## 6.4. Procedure

The aim of the study was explained to participants. They were given free trials to use the laser tracker device to reduce the learning effect of the modality. After the learning trials, they undertook actual trials of study. Participants were instructed to drive without deviating from the lane. While drivers were not instructed to maintain any speed limit, the software system capped the top speed at 60 Km/h. After each trial, participants were instructed to fill up NASA TLX and SUS forms. All trials were randomised to avoid the order effect.

## 6.5. Results

We have measured the following dependent variables

1. **Driving performance** is measured as
   a. **Mean deviation** from designated lane calculated according to Annex E of ISO 26022 standard.

    b. **Average speed** of driving, we investigated if the new modality significantly affected driving speed.

    c. **Standard Deviation of Steering Angle**, a large standard deviation means drivers made sharp turns for changing lanes.

2. **Pointing and Clicking performance** is measured as

    a. **Error in secondary task** as the number of wrong buttons selected compared to total number of selections.

    b. **Response time** as the time difference between the auditory cue and the time instant of the selection of the target button. This time duration adds up time to react to auditory cue, switch from primary to secondary task and the pointing and selection time in the secondary task.

3. **Cognitive load** measured as the NASA Task Load Index (TLX) score.

4. **Subjective preference** as measured as the System Usability Score (SUS).

In the following bar charts, heights of bars represent arithmetic mean, and the error bars represent standard deviation. We first undertook ANOVA and then compared means through pairwise t-tests after Bonferroni correction.

**Driving Performance:** The arithmetic means of the deviation and speed were compared across modalities as metrics of driving performance. The touch modality showed the least mean deviation in Figure 16. The mean deviation for laser tracker with mechanical switch was significantly (two-sample t-test: $p<0.05$) higher than that of laser tracker with eye gaze switch (ANOVA: $F_{(2,27)} = 222.9174$, $p < 0.01$). The average speed of driving was highest for touch interface, as seen in Figure 17. The average speed was not significantly different for laser tracker with mechanical switch and laser tracker with eye gaze switch, though the difference was less than 6 km/h.

**Pointing and Selection times:** The average selection time of target was lowest for the touch interface. It was significantly different for different conditions (ANOVA: $F_{(2,246)} = 8.0277$, $p < 0.01$). The touch input was significantly faster ($p<0.05$ at pairwise t-test after Bonferroni correction) than Laser pointer with mechanical switch, other two t-test results were not found significant. The difference between the average selection time of laser tracker with eye gaze switch and touch input was less than 100ms, as seen in Figure 18.

**Subjective Feedback:** The perceived mental workload was measured as the workload score from TLX. The TLX score was the least for the touch interface, as seen in Figure 19. It was also found through a two-sample t-test that the TLX score of the laser tracker and touch interfaces were not significantly different from each other.

The subjective preference was measured as the usability score from SUS. The SUS score was highest for the touch interface and was greater than 68 for all the three modalities. It infers that all systems are usable, as seen in Figure 20. It was also found through a two-sample t-test that the SUS scores of laser trackers were not significantly different from that of the touch interface.

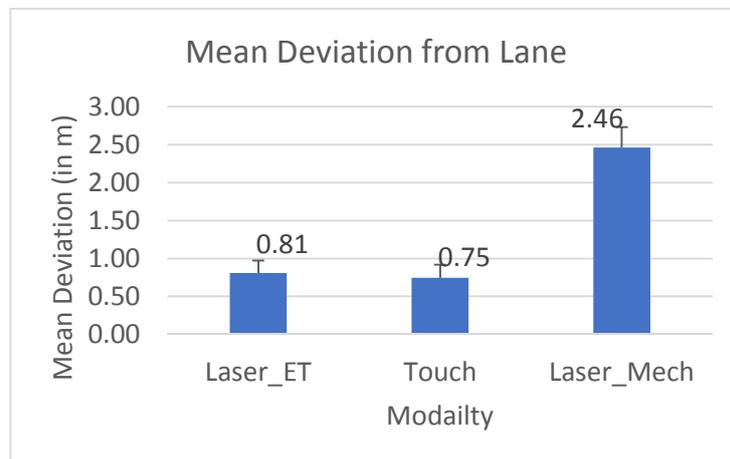

Figure 16. Comparing driving performance in terms of mean deviation from the lane

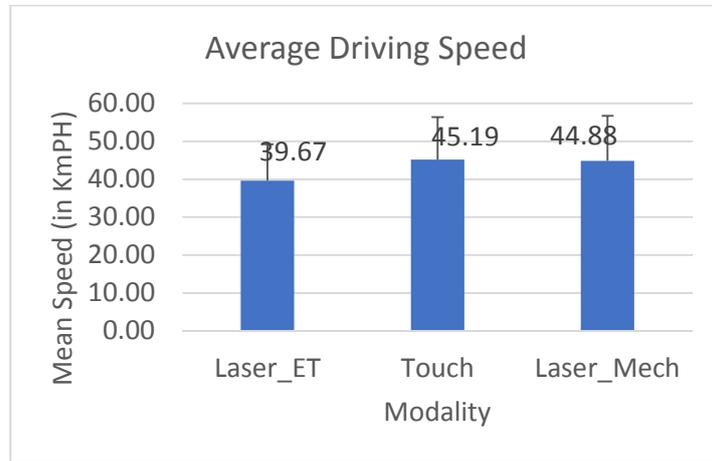

Figure 17. Comparing the driving performance in terms of the driving speed

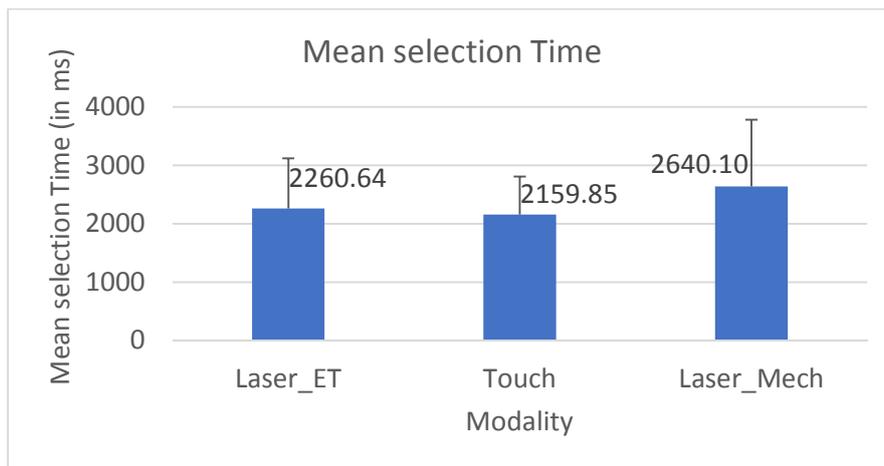

Figure 18. Comparing average selection time of the targets on the IVIS

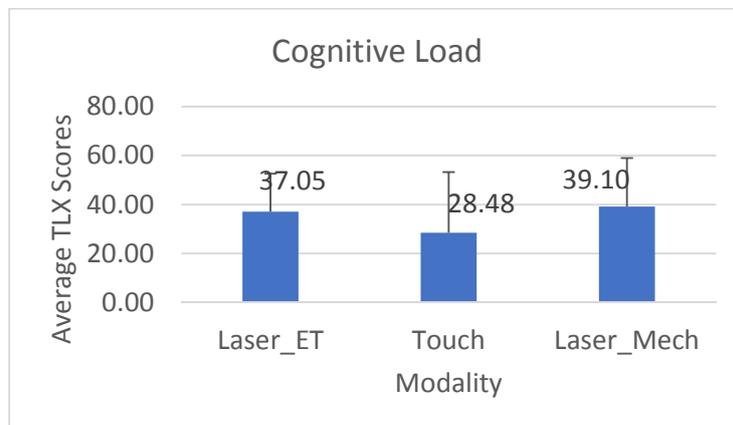

Figure 19. Comparing cognitive loads in terms of TLX scores

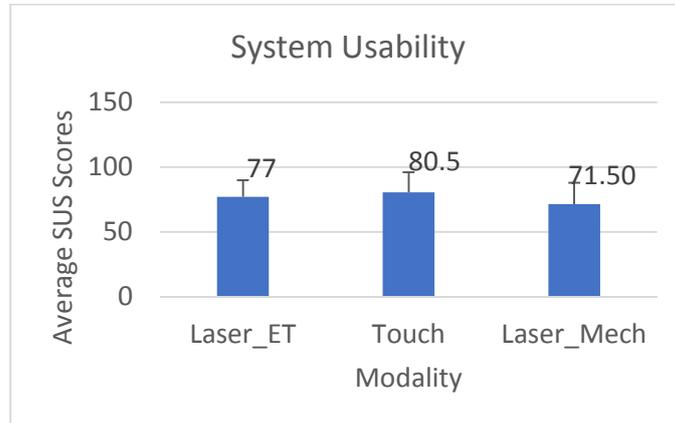

Figure 20. Comparing system preference in terms of SUS scores

## 6.6.     Discussion

This study demonstrated that the laser tracker interface could be used for performing a secondary task alternative to conventional button and touch interfaces. However, the touch interface was the fastest to use. The mean deviation plot (Figure 16) may indicate that participants were less distracted by the selection process using laser tracker with eye gaze switch than by using laser tracker with mechanical switch. The average speed plot may indicate that participants slowed down by 6km/h while performing a secondary task using laser trackers. Though participants took less time for selecting the intended targets using touch interface, the difference in selection time between modalities was less than 100ms, which infers that the performance of laser tracker interface is still on par with the touch interface. We expected the average selection time of the target to be significantly less for laser tracker than touch modality, but the experimental results did not support the hypothesis. One reason might be that users were not trained to operate a laser tracking device and they have already been well trained with touch modality. In our experiment design, we did not include fatigue as a parameter as it could give how comfortable or tired one felt while performing tasks using both laser and touch modalities.

However, one disadvantage of our system is that it is based on wearable technology and can be considered intrusive by users. It also requires the user to stretch his arm, although the duration and amplitude of stretching can be reduced compared to the touchscreen system as viewed in this video [Link]. Our recent work explored eye gaze-controlled HUD, which does not require taking hands off from the steering wheel [Prabhakar 2019]. However, our laser tracker technology remains relevant for HDD and can be made operational using a low-cost webcam-based eye gaze tracker. To address the implementation constraints of our laser tracker system inside a real car, we conducted a user study to evaluate the performance of our system inside a real car, as discussed in the following section.

# 7.    User study inside car

The laser tracker with eye gaze switch was set up inside a non-moving Toyota Etios car, as shown in Figure 21. The study aimed to organise various components of the system like camera, wearable device, touch sensor for steering wheel – all within an actual vehicle and externally validate the system's performance in terms of average selection time using a standard ISO pointing task.

## 7.1.    Participants

Data were collected from 10 participants (4 female, 6 male), with an average age of 29 years. Participants sat at the front passenger's seat.

## 7.2.    Material

A laptop with a matte finish display was set up above the gearbox such that the display was positioned in front of the existing infotainment system of the car. The camera was placed between the front seats such that the display was within its field of view without occlusion.

## 7.3. Design

The eye gaze tracker was calibrated and placed above the display of the laptop. The eye gaze tracker was tilted towards the participant such that it could track the eyes of the participant. Participants undertook the standard ISO 9241 pointing task using laser tracker with eye gaze switch. They were instructed to select a target button highlighted in white colour among a set of 5 blue-coloured buttons. Each button was $70 \times 70$ pixels in size. The difference in timestamps between the appearance of the target and its selection was recorded. We used same set of materials as described in section 6.

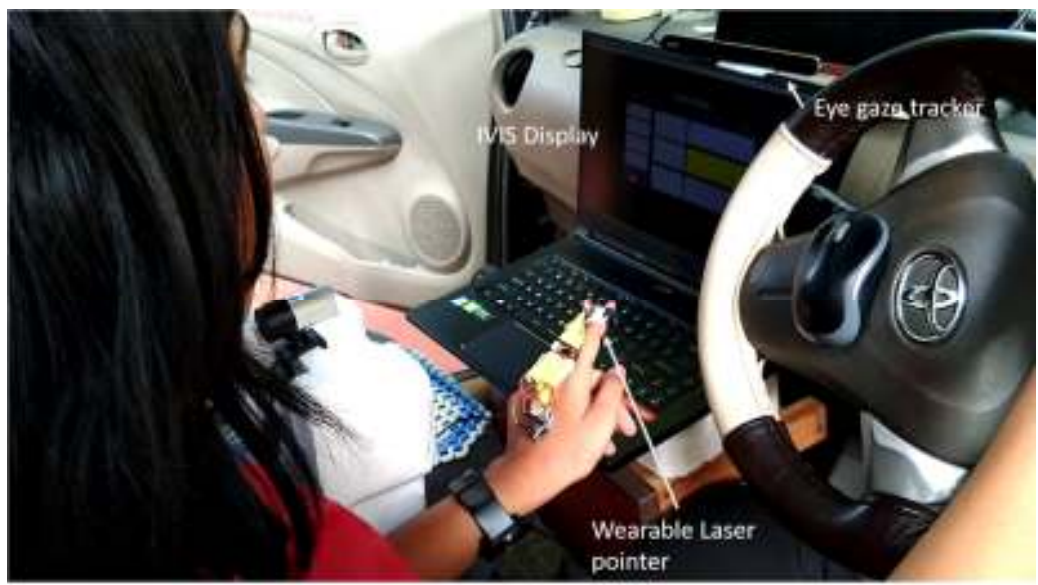

Figure 21. Participant performing secondary task using laser tracker with eye gaze switch inside car

## 7.4. Procedure

Before trial, participants were briefed about the set up and task. Each participant initially undertook a trial run followed by the actual task. Each participant continued the ISO pointing task for at least 2 minutes.

## 7.5.    Results

We recorded 348 pointing tasks from all participants, out of which 15 entries were removed as outliers as the values were outside outer fence ($3^{rd}$ Quartile $\pm$ 3 $\times$ Interquartile Range). On average, no participant exceeded 2 seconds to complete pointing tasks, as shown in

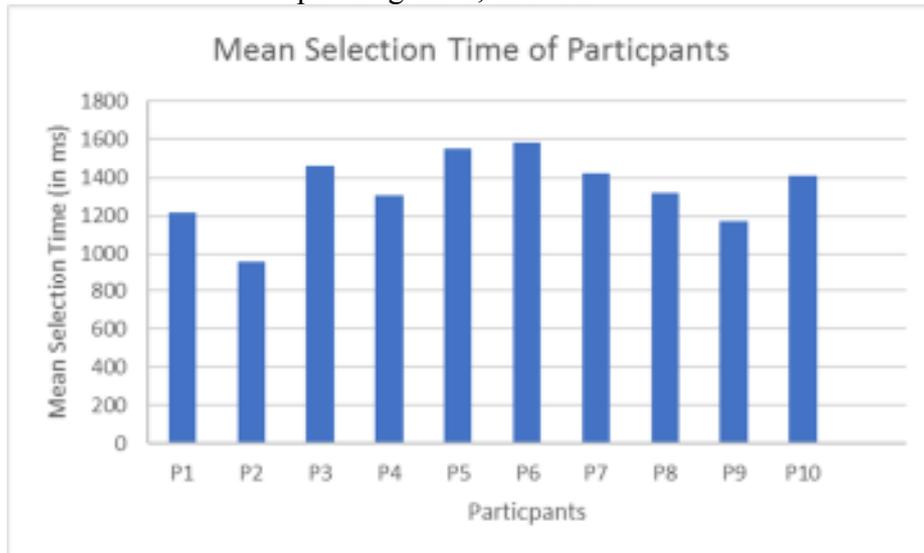

a. Participant wise analysis

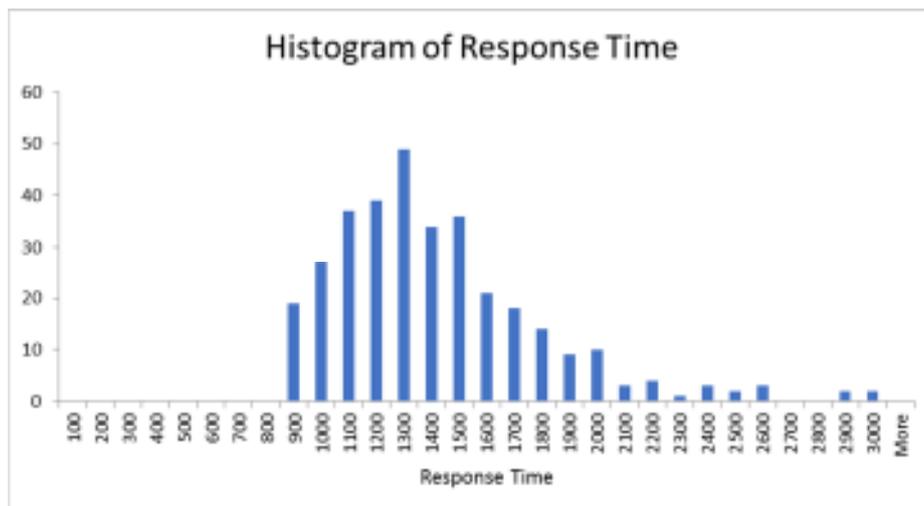

b. Histogram analysis

Figure 22a. The mean, median, and standard deviation of selection times were 1323ms, 1282ms and 308ms, respectively. A histogram in

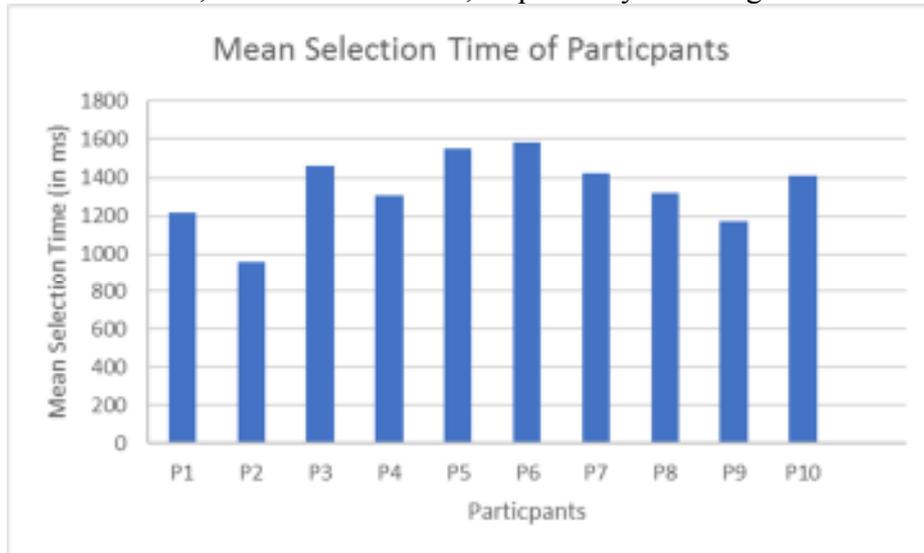

c. Participant wise analysis

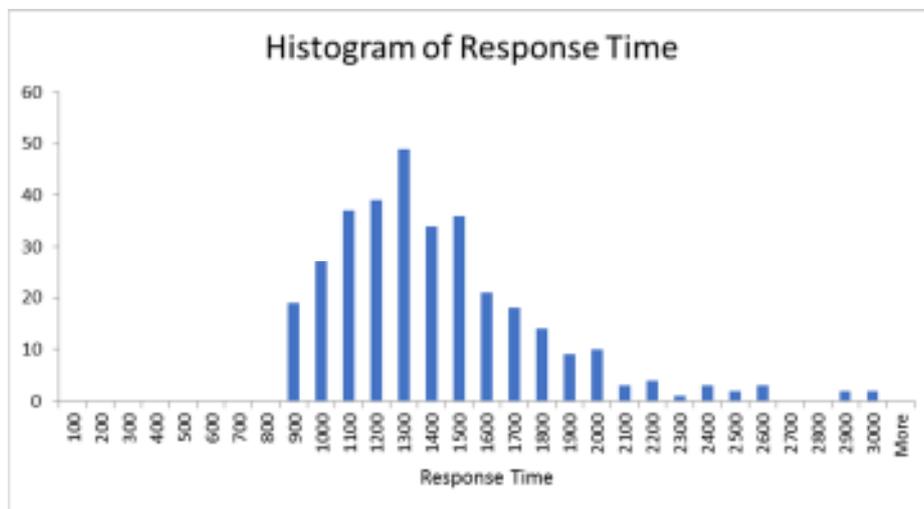

d. Histogram analysis

Figure 22b further showed that 94% of pointing tasks were completed in less than 2 seconds.

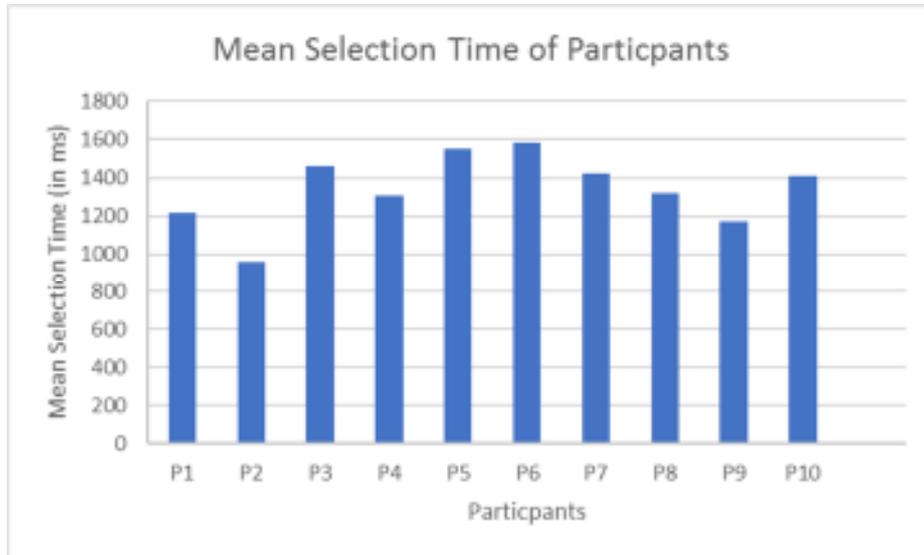

e. Participant wise analysis

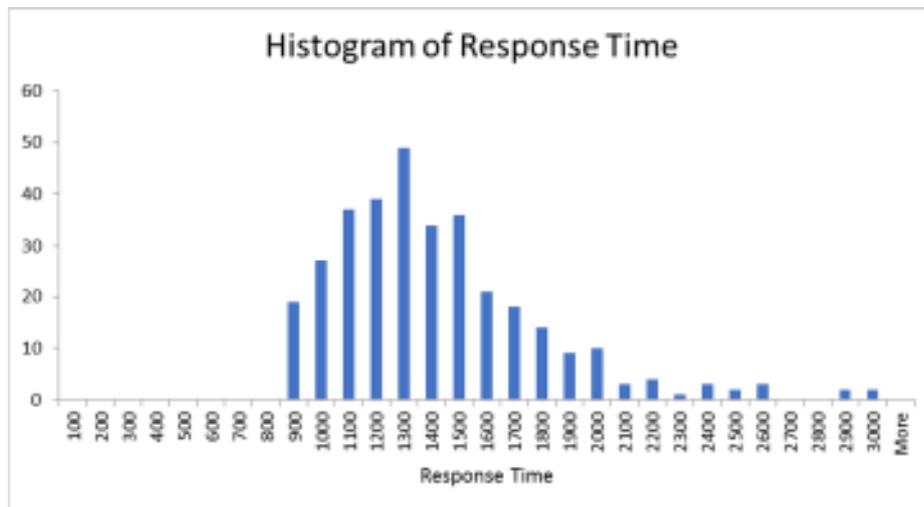

f. Histogram analysis

Figure 22. Pointing and selection time analysis for in-car study

## 7.6. Discussion

Initially, the laser tracker system was compared with IR and IMU-based hand movement trackers. Participants could undertake pointing and selection tasks faster using the laser tracker-based system. The laser tracker and eye gaze tracker were evaluated for their efficacy in dynamic lighting conditions in both laboratory and in-car environments. The laser tracker with eye gaze switch was integrated into a driving simulator and evaluated for its efficacy in completing pointing tasks in IVIS while driving. It showed promising results to investigate the laser tracker system further inside a real car. As our laser tracker system requires a camera to be mounted such that it faces the IVIS display, it was challenging to integrate the system in a real car and calibrate the camera for optimal field of view and exposure values. We conducted an in-car trial to evaluate the efficacy of the system despite the limited space. The performance of laser tracker with eye gaze switch inside the car was promising. Our laser tracker system took an average selection time of less than 2s, which is one of the main reasons to select this technology in real driving for operating secondary tasks without distracting from driving task. Participants felt easy to operate our system inside the car as they did not require to stretch their arms to make physical contact with the screen for interaction. Due to the quick response times and non-contact interaction, this system widens the scope to rethink new automotive user interfaces. In other intelligent virtual touch-based systems [Ahmad 2014; Biswas 2017], the predicted screen element will not be obvious to the user, and the IVIS software has to be modified to give visual feedback to the user about the predicted screen element. However, the laser tracker system will always show visual feedback with a laser spot reducing the chances of wrong selection. Considering the potential of our laser tracker technology for HDD to be adopted by industries, we investigated the acceptance of our technology by the end-users (drivers) by using qualitative research methodology, as discussed in the following section.

## 8.    Qualitative study for user acceptance

This study explores the contributory factors of acceptance of the proposed wearable device in a car by analysing why and how professional drivers would accept the proposed device. It helped us to understand professional drivers' expectations and opinions that will help in improving the prototype. There exists a research gap in understanding how new technology will be accepted and adopted by professional drivers in a service-oriented economy.

There are several models for technology acceptance starting from the year 1974, like the Theory of Reasoned Action (TRA) Model, Technology Acceptance Model (TAM), The Unified Theory of Acceptance and Use of Technology (UTAUT) Model. These theories were developed keeping the broader context, information technology, and specific context in mind. It was found that these models were useful tools for assessing the plausibility of flourishment of introducing new technology. These models contributed to the knowledge of factors that facilitate acceptance and those that lead to declining acceptance. It was found that constructs like perceived usefulness, perceived comfort/pleasantness, attitude towards technology, user-friendliness, trust, purchasing power, perceived accuracy, and social influence are good predictors of acceptance of new technology [Schmidtler 2017, Spagnolli 2015]. Several studies were conducted to understand, explore, and evaluate various wearables, assistive technologies to support designers. As wearables are widely accepted in economically developed countries, it is crucial to understand factors contributing to the acceptance and adoption of new technology in economically developing countries [Debnath 2018]. Our study explored dynamics that contribute to the acceptance of wearable laser tracker-based HDD and emerging areas that need investigation. This study also helped us to understand how individuals respond to events or handle problems through action and interaction.

## 8.1. Participants

A total of 11 male Indian professional drivers between the age group of 28 to 58 years with a four-wheeler driving license voluntarily participated in the study. Participants spoke a diverse range of languages like Kannada, Tamil, Telugu, and English. Informed consent was obtained from all participants. Confidentiality and privacy were ensured to generate honest responses as much as possible. All participants but one owned a vehicle like Toyota Etios and Maruti Swift Desire that was no older than seven years. Questions about the history of accidents were asked to which they all said that minor accidents of the vehicle are unavoidable in the traffic of Bengaluru. All participants had prior experience of outstation driving. They ride a minimum of 100 km per day inside the local city. A few participants had the habit of smoking and suffered from health problems like blood pressure.

## 8.2.    Material

We used a semi-structured interview questionnaire for interviewing participants and recorded sessions using voice recorders. We used a driving simulator with a simulated dashboard and driving seat. The dashboard was set up as a Head Down Display (HDD) on the central stack, and the laser tracker system was integrated into the dashboard, as shown in Figure 23.

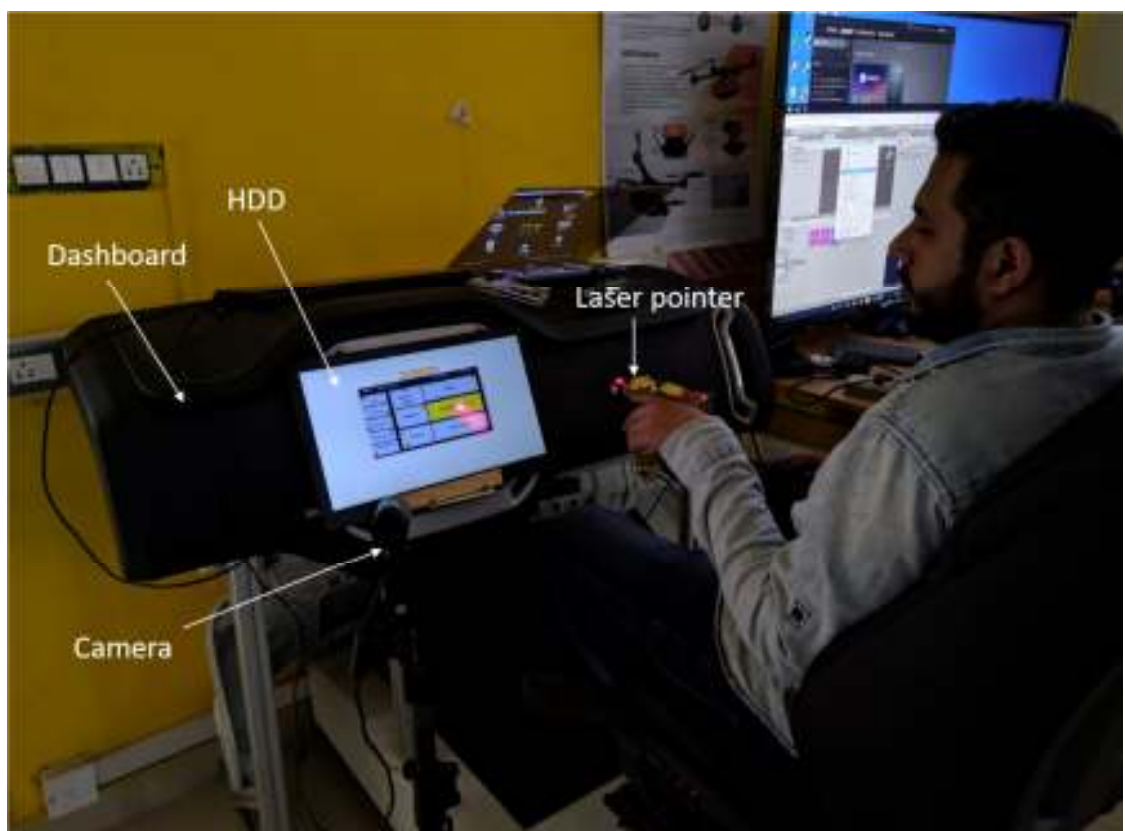

Figure 23. Screenshot of wearable laser tracker-based HDD

## 8.3.    Procedure

Participants were initially demonstrated the working of the system and then they used the system as long as they wanted for a maximum duration of 10 minutes. We did not collect any quantitative data from the trial and participants were allowed to ask question or comment while they were using the system.

An interview guide was developed, which consisted of open-ended questions that allowed participants to give a free-form answer. This interview guide was prepared after reviewing literature on the acceptance of technology and wearables. Participants were given time at the end of each interview to add different topics beyond the questionnaire that seemed important to them. This type of interview helped to maintain consistency over concepts that were covered in each interview. A research diary was maintained, consisting of participants' appointments, summaries of discussion, problems, dates, and important information that was key to the research. Face-to-face interviews were conducted. Participants were asked the language that they were most comfortable with and proceeded with interviews in that language. A translator was appointed to prevent the loss of information and meaning of data in the interview. Interviews were conducted for over three weeks. Interviews were held for 15 to 50 mins, with an average interview time of 20 mins. The grounded theory method [Corbin 2015] of analysis was used for a systematic, comparative, inductive, and interactive approach to analysis. Systematic comparisons and inquiry of concepts were integrated throughout the data collection and analysis. Theoretical sampling [Corbin 2015] was aimed to achieve.

## 8.4.    Data analysis

Glaser & Strauss method [Corbin 2015] was used for data analysis. It allows identifying general concepts, development of theoretical explanations and offers novel insights into human experiences. Questioning, making constant comparisons, thinking about various meanings of words, using the flip flop technique [Corbin 2015], looking at the language and emotions were used to analyse the data. Memos and diagrams were maintained from the beginning of the analysis. Theoretical sampling was ensured to understand the concepts elaborately and was repeated until saturation was attained. Open coding, axial coding, and selective coding were used for creating and connecting categories.

Transcripts were looked for sections or paragraphs. A section from the transcript was taken, and a concept is denoted to it. These concepts differed in levels of abstraction. The right words that conceptually best described the meaning of the data was chosen as concepts. These concepts were lower in the level of abstraction and were provided by participants. They were called "in-vivo codes" or open codes. Higher-level concepts were developed after the generation of open codes. They were called axial codes or categories and represented more abstract terms. Major themes were generated from the basic concepts (open codes). Open and axial codes were continued to be generated and developed from transcripts until sensitivity to meaning grew, which helped to form categories. Each category consisted of conditions, actions–interactions, and consequences to sort out and arrange concepts by asking questions and thinking in terms of possible linkages. Finally, a suitably extensive, eclectic, and abstract concept, which outlined the main ideas in a few words expressed in the study, was chosen. It was called the core category.

Properties and dimensions like the intensity, duration, range of the concept were noted. The data was referred again and was looked for further descriptions of the same construct in the same interview. The same construct was also examined in other interviews, and those descriptions were compared. The concepts, constructs, and their relationship became clearer with this process. What to look for in the upcoming interviews was noted from the memo. Memos were dated and marked along with the use of conceptual headings. The initial memos were descriptive, and as the analysis proceeded, it became more conceptual. Flexibility and free flow of thoughts were ensured throughout the memo-writing process. Memos were regularly sorted, reordered, and visualised through network diagrams to ensure appropriate linking of the various concepts.

## 8.5. Results

The open codes were collected from transcripts. The axial codes are generated from the comparison of open codes using memos and diagrams. Figure 24 illustrates the axial codes and their influence on acceptance. User Experience, desire for control, technology anxiety, social influence, purchasing power, age, driving experience, and vehicle ownership are the generated axial codes from this study. Age influences user experience, desire for control, technology anxiety, social influence, and purchasing power. Driving experience influences user experience, desire for control, technology anxiety, and purchasing power. Ownership of vehicles influences purchasing power and desire for control. User experience is influenced by likeability, learnability, ease of use, and perceived usefulness. All these twelve axial codes contribute to the development of acceptance as the core category.

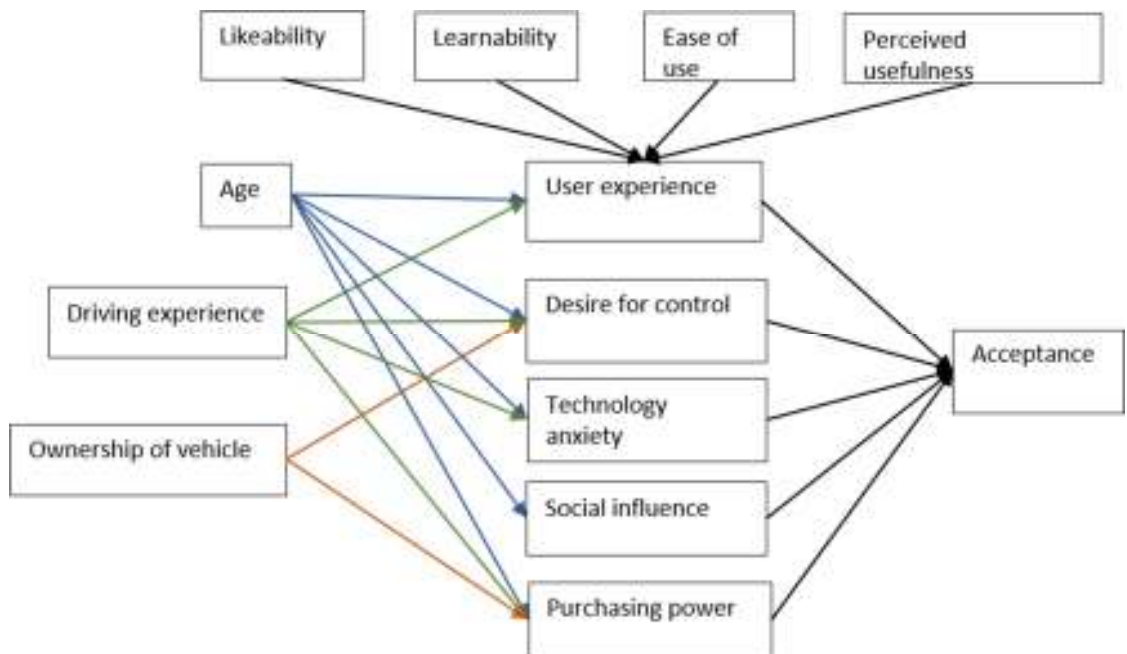

Figure 24. Acceptance model of wearable laser tracker-based HDD

## 8.6.    Discussion

We noted that a significant number of existing user interfaces in automotive have evolved without following a user-centric design process [Schnelle 2019]. Though researchers found innovative ways to establish such natural interactions, which are effectively invisible in cars and reduce learnability, these innovations often do not reach the market. It is due to the non-user-centric design approach, which does not involve users in every stage of design and development of interactive technologies for automotive. In this paper, we followed a user-centric design process and undertook quantitative and qualitative analysis to improve the invention. In the following paragraphs, we summarised end users' expectations for a new interactive device.

**User Experience:** After using the prototype for the first time, participants felt it was swift, easy, compact, useful, lightweight, and simple to use. It was found easy to learn as there was no hand movement required, and the task could be established using a single finger.

**Desire for Control:** The desire for control is a personality trait defined as the extent to which individuals are generally motivated to feel as if they are in control of events in their lives [Amoura 2014]. Passengers using the infotainment system sometimes spoil the touchscreen, which drivers need to repair themselves. Wearable laser tracker-based HDD facilitates prevention of in-vehicle damages and adds to the positive mood of drivers.

**Technology Anxiety:** Anxiety refers to a negative valence that is accompanied by a sense of uncontrollability and focused on a possible future threat or other potentially negative events. Technology-induced anxiety (often called computer anxiety) elicits negative emotions, apprehension, and fear that arise because of the introduction of new technology [Nordhoff 2018]. Wearable laser tracker-based HDD is easy to learn and has familiarised concept of laser pointing which generated less anxiety among drivers.

**Social Influence** refers to how individuals change their ideas and actions to meet requirements of a social group, perceived authority, social role, or a minority within a group wielding influence over the majority [Schmidtler 2017]. Social influences here are colleagues, friends, relatives who own vehicles. Local garage and accessories outlet sellers also influence the decision-making process of acceptance of new technology.

**Purchasing Power** is defined as the amount of money that a person or group has available to spend [Merriam-Webster 2020, Hossain 2017]. Participants speculated the price of the product to be in the range of five thousand to forty thousand INR. It was the estimated price that they were willing to pay for a wearable laser tracker-based HDD. The purchase depended on multiple factors like size of display, car loan, available features, maintenance cost, earnings, and the subsidy provided by the organisation. It was suggested that such systems should be pre-installed and affordable. According to them, the showroom pricing and the accessories pricing are different, and they like to go forward with the more affordable ones. If the maintenance cost is less for the product, then it is an added benefit.

**In summary**, drivers reported higher response time and malfunctioning of the touchscreen at crucial times like picking up passengers, as the main problems with existing touchscreen-based IVIS. All 11 professional drivers were able to use the laser tracker-based virtual touch system. They thought that the system is easy to use and learn and can increase the durability of the existing touchscreen by reducing the number of physical touches. They wanted the system built into the car dashboard and also estimated a price range for the device. They identified lower response time as a success criterion for the system.

## 8.7.    Value Addition

Researchers explored wearable technology mostly for physiological parameters measurement, while virtual touch interfaces mostly focus on hand tracking and movement trajectory prediction.  This research explored wearable systems as a virtual touch system for direct manipulation interface and reported a set of case studies in the automotive environment. Laser pointer is a common technology used in consumer electronics and less explored in interaction technologies. This research explored virtual touch with a laser pointer and found that it competed with the existing touchscreen interface in terms of the selection time for operating IVIS and driving performance in both simulation and in-car environments. The user studies involving virtual touch and eye gaze tracker in different lighting conditions will be useful for researchers in Human-Computer / Machine Interaction (HCI/HMI). The laser tracker-based pointing system can be useful for interacting with large screen displays or digital television [Biswas 2017]. In terms of deployment, the laser tracker can be integrated with smart key of cars, while any general-purpose camera can work as the eye-tracking switch as the system does not require less than $1.6^0$ of accuracy for activating the switch.

Technology acceptance by consumers is a key aspect of any new automotive technologies [Adell, 2009]. We investigated factors influencing the acceptance of the proposed technology, which can help in further modifying the existing designs of prototypes for a better user experience. The qualitative study found attributes contributing to user acceptance of new interaction technologies in cars.

## 9.    Conclusions

This paper presented a new wearable pointing device for operating IVIS inside a car through virtual touch. The initial study compared three different technologies for virtual touch systems. The laser tracker was evaluated in automotive environment as it showed the best performance in terms of accuracy and selection time in the initial comparative study. We compared a mechanical switch and an eye gaze switch for selecting the target with laser tracker. We integrated the system with the IVIS of a driving simulator for undertaking secondary tasks while driving. The performance index, cognitive load, and system usability were evaluated. The laser tracker with eye gaze switch performed better than the laser tracker with a mechanical switch in selection time and driving performance. It was found that the performance of laser tracker-based systems did not significantly deviate from that of a touch-input controlled system. A study evaluated the performance of the laser tracker system inside a non-moving car. Results of this study were promising as the average selection time was less than 2 seconds. Finally, a qualitative study found that the user experience, desire for control, technology anxiety, social influence, and purchasing power influenced user acceptance of our technology. Future work is planned to improve the ergonomics of the system for better usability and reduced distraction.